# Onset condition of the ramified-jet regime: an experimental and theoretical study

Bor Zupan[1], Jernej Kušar[1], Božidar Šarler[1]

[1]Laboratory for Fluid Dynamics and Thermodynamics, Faculty of Mechanical Engineering, University of Ljubljana, Ljubljana, Slovenia

**Correspondence**: bozidar.sarler@fs.uni-lj.si

**Abstract**: This study experimentally and theoretically investigates the electrically induced transition of a gas-focused liquid jet into a rim-emitting sheet. Ultra-high-speed imaging was used to examine the effects of applied voltage, gas and liquid flow rates, and liquid properties on sheet formation and breakup. Fully developed, quasi-stable sheets formed with ethanol, whereas isopropanol exhibited only initial sheet formation. Increasing voltage promoted sheet stretching, finer and more numerous rim jets, and a narrower spray. Higher gas flow rates shifted the transition towards the nozzle and produced finer structures, while higher liquid flow rates generated larger sheets with thicker rims. A first-order theoretical model assumes an equipotential interface and neglects finite charge relaxation, hydrodynamic perturbations, and nonlinear evolution. It balances destabilizing Maxwell-pressure perturbations against restoring capillary-pressure perturbations on a conducting cylindrical jet. The model identifies the elliptic azimuthal mode, $m = 2$, as the first unstable mode leading to sheet formation and predicts a critical electric Bond number of 3. Predicted onset voltages agree with measurements within 20%, and the threshold was consistent with published ramified-electrospray observations. The agreement with experiments supports electrocapillary symmetry breaking as the mechanism of onset and provides a compact, physically grounded criterion for predicting the transition from electrified jets to rim-emitting sheets.

## 1 Introduction

Electrohydrodynamic atomization is governed by the competition among liquid inertia, viscosity, surface tension, charge transport, and Maxwell stresses. Depending on the operating conditions and liquid properties, this competition produces numerous spraying modes, including dripping, spindle, cone-jet, oscillating-jet, multijet, and ramified-jet regimes[1]. Most previous studies have concentrated on the cone-jet regime because it generates small, highly charged droplets with a relatively narrow size distribution. At larger flow rates or stronger electrification, however, a cylindrical jet can undergo a non-axisymmetric deformation in which it broadens into a sheet or ribbon and emits secondary jets from its periphery[2].

Liquid sheets can also be produced by purely hydrodynamic means. Impinging liquid jets form flat sheets through momentum redirection, while gas-accelerated nozzles use aerodynamic stresses to generate thin, alternating orthogonal sheet structure. Such sheets provide short and comparatively uniform optical path lengths and have therefore been developed as sample targets for ultrafast spectroscopy, X-ray scattering, and serial femtosecond crystallography at X-ray free-electron lasers[3–6]. In these systems, sheet formation is imposed by the nozzle geometry and flow arrangement. The sheet considered here is fundamentally different: it develops spontaneously through electrically driven symmetry breaking of an initially cylindrical, gas-focused jet.

This electrohydrodynamic morphology was first reported by Huebner[2], who observed that strongly electrified isopropyl-alcohol jets developed a fan-like configuration with numerous secondary jets emerging from the fan edge. The behavior was attributed to the low surface tension of the alcohol, which reduced the capillary resistance to electrically driven surface growth. Jaworek and Krupa[1] subsequently classified related structures as the ramified-jet mode, characterized by a continuous mother jet producing dynamically changing lateral sub-jets. Nikitopoulos and Kelly[7] documented the collapse of a charged cylindrical oil jet into a thick-rimmed ribbon. They proposed that charge migration toward the extremities of a distorted cross-section stretches the jet into a sheet, after which the charge-rich rims become unstable and emit secondary filaments.

Recent experiments have established ramification as a reproducible electrospray regime. Rajabi et al.[8] observed a progression from varicose and kink instabilities to ramified jets and, at higher electro-Weber numbers, to sheets surrounded by multiple branches. Xu et al.[9] obtained stable ramified sheets using HFE-7100 blended with ethanol and distinguished pudgy- and lanky-ramified configurations. They quantified the sheet geometry and droplet distributions and constructed an empirical regime map based on the Reynolds and electric Bond numbers, while noting that a governing equation for the transition was still required. Wang et al.[10] similarly documented the evolution from varicose breakup through whipping and whipping-assisted bifurcation to ramified instability in high-flow-rate ethanol electrosprays.

Theoretical studies of idealized electrified liquid cylinders have shown that sufficiently strong electric stresses can destabilize non-axisymmetric azimuthal modes and produce elliptic deformation or jet splitting[11–13]. These studies establish the possibility of electrically driven cross-sectional symmetry breaking. They were not, however, formulated as experimentally applicable onset models for ramified sheet formation in an electro-flow-focused or electrospray configuration. Conversely, previous studies of ramified electrosprays have primarily provided morphological classifications, empirical regime maps, and qualitative arguments about charge migration. The remaining question is therefore not the identification of the ramified regime, but the prediction of its onset.

In the present work, we realize a ramified jet-to-sheet transition in an electro-flow-focusing system[14–17]. A liquid jet is aerodynamically focused by a gas-dynamic virtual nozzle and subjected to an axial electric field. Above a critical applied potential, the approximately circular jet loses its rotational symmetry and expands into a quasi-steady sheet bounded by pronounced rims. Numerous secondary jets emerge from these rims and break into droplets spanning a broad size range. We characterize the effects of applied voltage and gas and liquid flow rates on the formation and breakup of this rim-emitting sheet. To explain the transition, we develop a local linear onset model for an electrically stressed cylindrical jet. The interface is decomposed into azimuthal perturbation modes, and the destabilizing Maxwell-pressure perturbation is balanced against the capillary restoring pressure. Relating the surface field to the applied voltage through a geometry-dependent field-enhancement factor provides a directly testable expression for the critical voltage. This establishes the first onset model for the ramified-jet sheet formation.

## 2 Experiment

### *2.1 Experimental methods*

The experimental electro-flow focusing setup is shown schematically in Figure 1. An insulating electrical enclosure (A) protects experimental equipment from high voltage. Inside the enclosure, a triaxial translation stage (Thorlabs, US) with remotely controlled vertical movement precisely positions the GDVN (B), so it is concentrically aligned with the 2 mm orifice of the downstream electrode (C). Below the electrode is the impact plate (D), which collects the charged liquid sample. Glass viewports (E) are used for the illumination of the jet via incoherent light source (F) (Thorlabs, US) and visualization via a modular long distance working microscope (G) (Edmund Optics, US) and a monochromatic ultra-high speed camera (H) (Specialised Imaging, UK). A laboratory source (I) supplying dry air is connected to a thermal mass flow meter-controller pair (J) (Bronkhorst, NL). The exact thermodynamic state of the air is measured using a pressure (P) meter (First Sensor, DE) and a temperature (T) meter (Greisinger, DE) inside the equalization chamber (K). Both signals are collected with a measurement card (O) (National Instruments, US). Signals from the measurement card, flow meter and camera are collected by a computer (S). The liquid flow rate is ensured by a syringe pump (L) (Harvard Apparatus, US), while the electric potential is supplied using a high-voltage (HV) power source (M) (FuG, DE). The piezo-actuated vertical stage of the GDVN is positioned using a remote controller (R) (Thorlabs, US). The whole setup is positioned on an optical table (N) (Thorlabs, US).

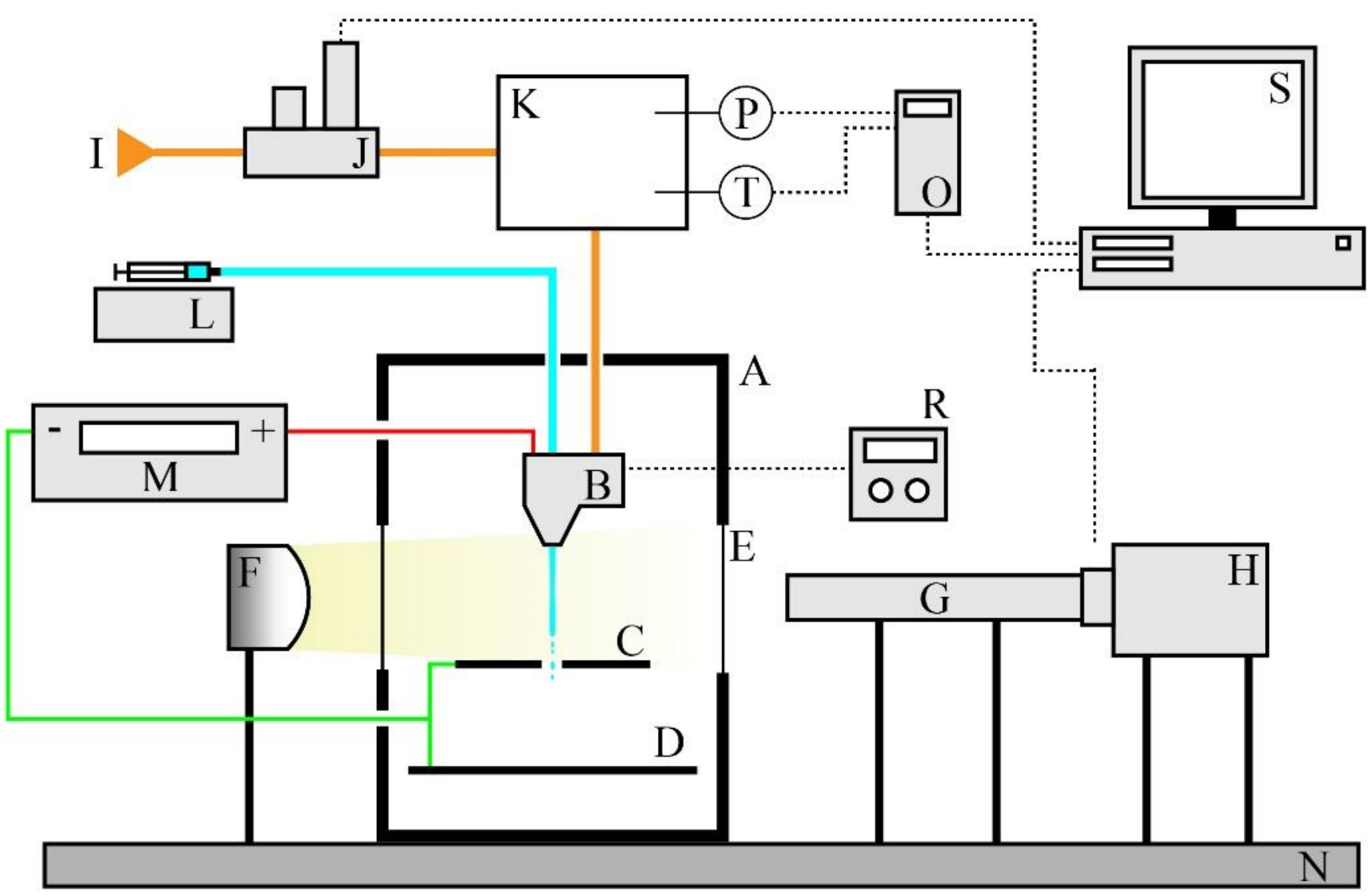

Figure 1: The experimental setup. Dotted lines represent signal transfer lines.

We used a standard GDVN with the liquid capillary diameter of 616 μm, outer orifice diameter of 960 μm, and capillary to orifice separation of 950 μm. The body of the nozzle was 3D printed with an SLA printer (Kudo3D, US) at 15x15x10 µm voxel resolution, cleaned with isopropanol, and cured under UV light. To ensure direct boundary mapping between the experimental and analytical approaches, we used a metallic liquid capillary inserted into a printed channel in the nozzle body. The capillary had an outer diameter of 794 µm and was directly connected to the HV supply.

A single experimental run proceeds as follows. The liquid and gas flow rates and electric field are initialized, and a 10-second stabilization period is provided to allow the quasi-steady jet to develop. During the measurement, pressure and temperature are collected at 100 Hz for 10 seconds and then time-averaged. The video frames are collected at 500 kHz, yielding 180 frames at 924x768 pixels with a pixel scale of 11.3 µm/px.

The focused liquids were ethanol (density 805 kg/m$^3$, viscosity 1.2 mPa s, surface tension 22.1 mN/m, relative electric permittivity 25, and electric conductivity 10$^{-8}$ S/m) and isopropanol (density 786 kg/m$^3$, viscosity 2.4 mPa s, surface tension 23.0 mN/m, relative electric permittivity 18, and electric conductivity 10$^{-8}$ S/m) at ambient conditions of 21 degrees centigrade.

*2.2 Sheet formation and breakup*

The sheet was formed as follows. First, the gas flow was established to clear the nozzle gas channel of any residual liquid droplets. Secondly, the liquid flow rate was initialized, resulting in either a steady-flow focused jet or a dripping/whipping regime. The electric potential was initialized at 6 kV, where a broad stable jetting space has been previously observed, which stabilized and/or formed the liquid jet. From there, the electric potential was gradually increased until the sheet was established, as shown in Figure 2.

The sheet remains quasi-stable in both time and space, while the secondary rim jets grow as they propagate downstream. These jets exhibit Taylor cone-like protrusions attached to the rim, which occasionally detach and travel downstream, as previously reported for microscale EFF jets[14]. The rim jets break up via the Rayleigh–Plateau instability, producing droplets of varying sizes. Droplet size depends on the position of the rim jet: thinner jets form near the beginning of the rim and generate smaller droplets. The rim closes without disrupting the sheet, unlike the behavior sometimes observed in gas-focused sheets[4], and no stable secondary sheet forms in the perpendicular direction. Downstream of the sheet, the collapsed rims form a highly unstable jet which continues to emit secondary jets until its final breakup.

The smallest droplets, produced by the first set of rim jets, are initially emitted approximately perpendicular to the primary flow direction. As they travel downstream, their trajectories bend toward the grounded electrode under the action of the electric field. These droplets reach higher velocities than the larger droplets because of their greater charge-to-mass ratio. The droplet mass scales as $m \sim R^3$, whereas the charge of a droplet near the Rayleigh limit scales as $q_R \sim R^{3/2}$, where $R$ is the droplet radius. Consequently, $q_R/m \sim R^{-3/2}$ , and the electrical acceleration, ($a_e = qE/m$), increases strongly with decreasing droplet size.

The breakup of the sheet ultimately produces a heterogeneous droplet-size distribution, with droplet size increasing toward the centerline. The largest droplets are comparable in diameter to the undeformed primary jet, whereas the smallest droplets fall below the optical resolution of our imaging system. The smallest droplets that can be measured reliably are approximately 10 µm in diameter; however, these do not originate from the earliest-formed rim jets, which appear to generate even finer droplets. This observation suggests that the smallest droplets may be on the order of 1 µm or less. These observations suggest that a millimeter-scale GDVN nozzle can generate micrometer-scale or smaller droplets from a low-electrical-conductivity liquid, such as ethanol. Such dimensions are typically associated with highly conductive electrosprays (using the same size liquid capillary) operating in the stable cone–jet regime and, to the best of our knowledge, have not previously been reported for flow-focusing systems at this scale.

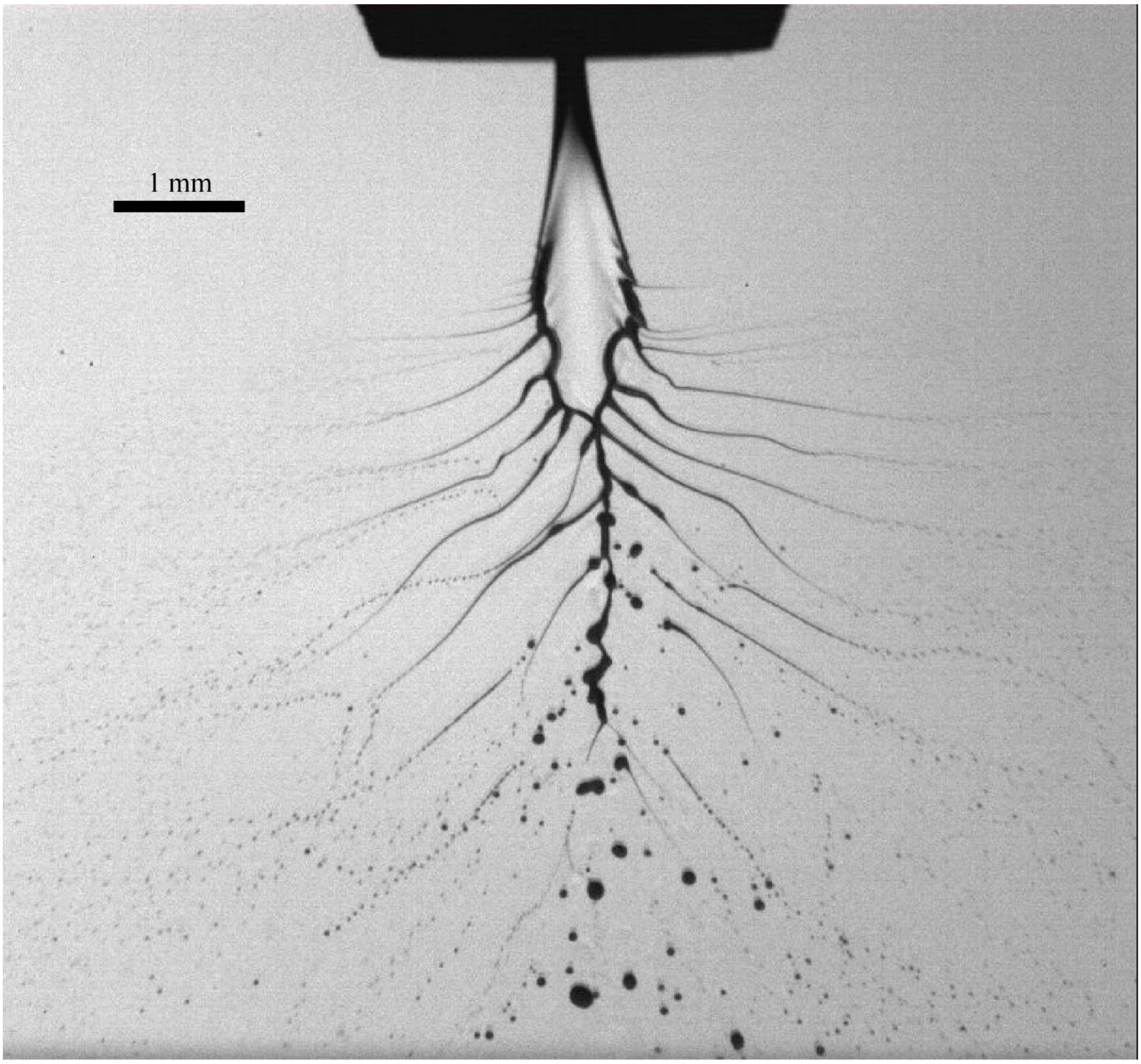

Figure 2: The rim-emitting electro sheet formed at 50 $l_n$/h air flow, 2 ml/min ethanol flow and 9 kV electric potential. Captured at 200,000 FPS.

### *2.3 Influence of the electric field*

Increasing the applied voltage strengthens the electric field and, consequently, the Maxwell stress acting on the charged liquid interface. The enhanced electrical stress stretches and thins the sheet while promoting the formation of secondary jets along its rims. After emission, these jets and the resulting droplets are increasingly deflected toward the grounded electrode, as evidenced by the progressive downstream alignment and narrowing of the spray envelope shown in Figure 3. This effect is particularly pronounced for the finest jets and droplets because of their larger charge-to-mass ratios.

Although the overall sheet topology persists throughout the investigated voltage range, the downstream jet remains highly unsteady, and its instantaneous oscillation amplitude and breakup position do not exhibit a strictly monotonic dependence on voltage. Furthermore, at higher voltages, the enhanced stretching and thinning of the sheet, together with the formation of more numerous rim jets, redistribute the available liquid flow among more structures. Consequently, the volumetric flow rate supplied to each rim jet decreases, resulting in smaller rim-jet diameters.

The reduced attenuation of the backlight through the sheet at higher applied potentials further suggests that the sheet becomes progressively thinner, although this observation remains qualitative in the absence of an optical-thickness calibration.

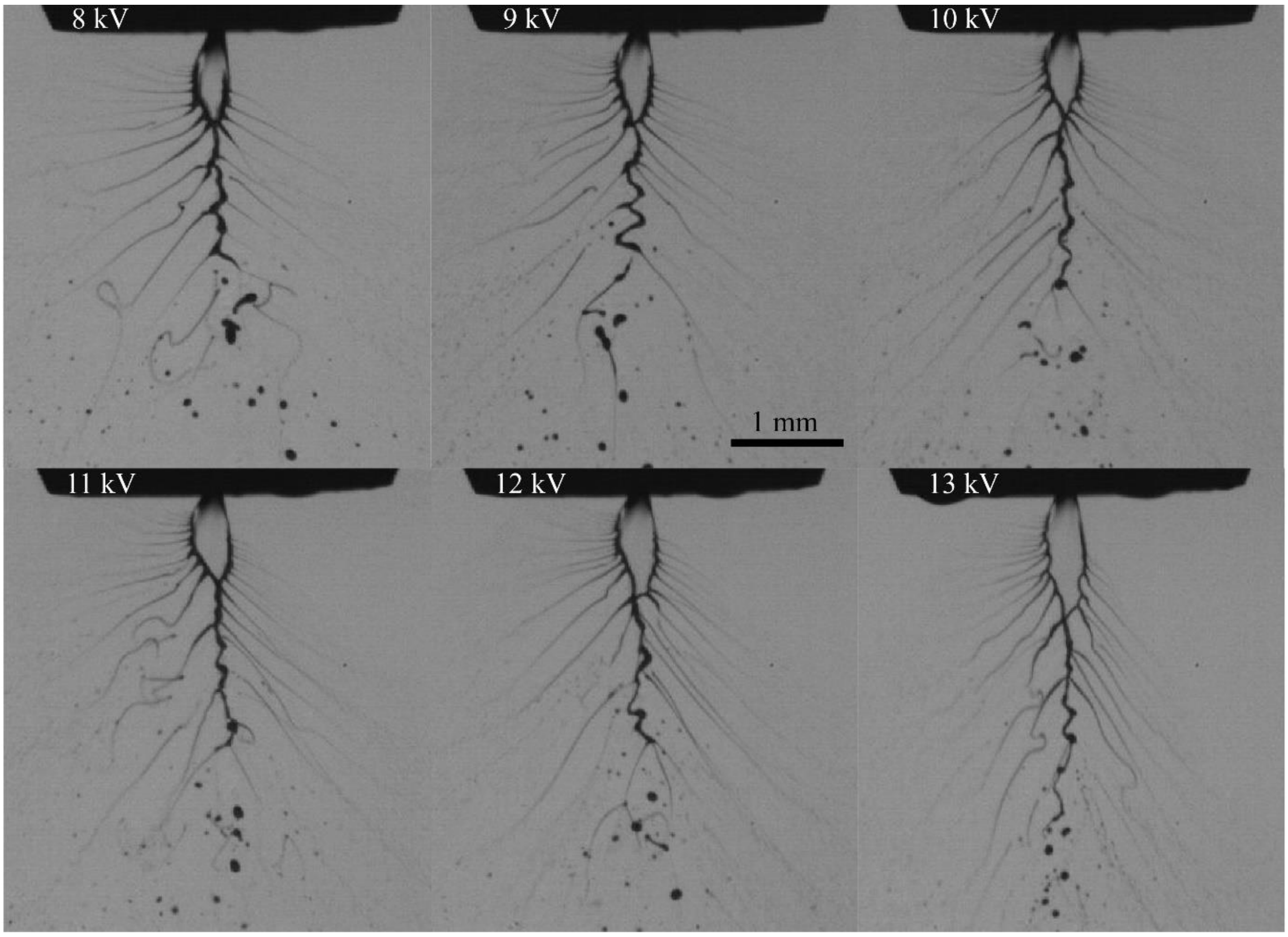

Figure 3: Influence of the electric field on the rim-emitting electro sheet. All cases were filmed with 50 $l_n$/h air flow and 1 ml/min ethanol flow at 500,000 FPS.

*2.4 Influence of the liquid and gas flow rates*

Increasing the gas flow rate (see Figure 4) strengthens the aerodynamic stresses acting on the liquid, causing the jet-to-sheet transition to occur closer to the nozzle and promoting greater stretching and thinning of the sheet. The resulting rims and secondary jets also become finer. By contrast, increasing the liquid flow rate increases the liquid inertia and the volume supplied to the sheet, producing a longer, wider, and optically denser structure with thicker rims and a more pronounced central jet.

At the lowest gas flow rate (see Figure 4), the focusing stress is insufficient to accommodate the highest liquid flow, resulting in an irregular ligamentary structure rather than a coherent sheet. These observations indicate that sheet formation is also controlled by the gas-to-liquid momentum balance: higher gas-to-liquid ratios favor thin, strongly focused sheets, whereas lower ratios produce larger and thicker sheets and, eventually, unstable breakup.

*2.5 Influence of the liquid properties*

We have successfully attained a fully developed rim-emitting electro sheet regime using ethanol. Any attempts using other common solvents with higher surface tension, like water (72.8 mN/m) or ethylene glycol (48.0 mN/m) have been unsuccessful at this scale.

Although isopropanol has a surface tension comparable to that of ethanol (23.1 mN/m), its viscosity is approximately twice as high (2.4 mPa s). Initial sheet formation was observed at higher applied voltage and liquid flow rate than that of ethanol, but the jet disintegrated before a fully developed sheet could form.

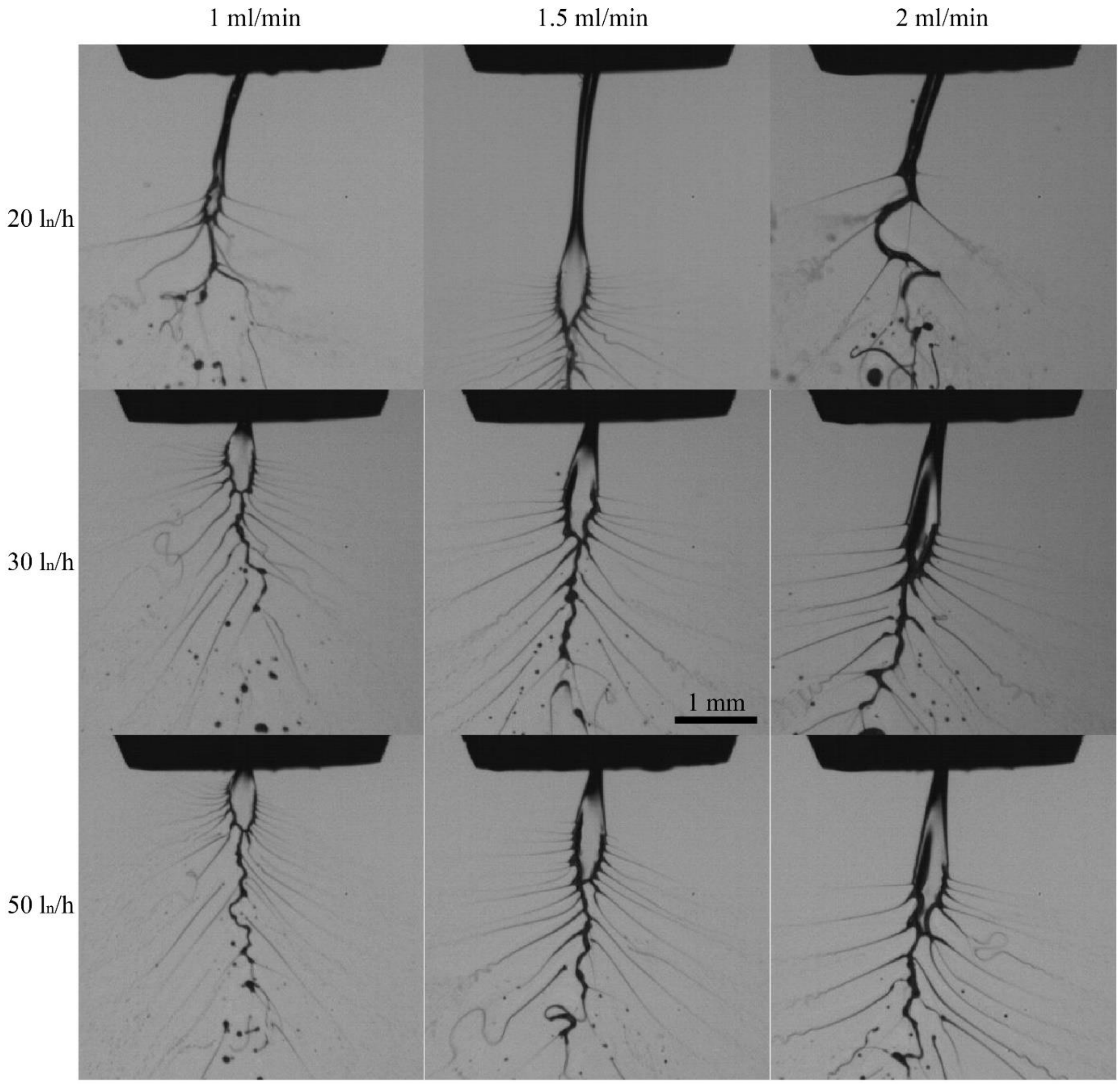

Figure 4: Influence of the gas (rows) and liquid (columns) flow rate on the rim-emitting electro sheet. All cases were filmed at 10 kV and 500,000 FPS using ethanol.

## 3 Theoretical analysis

We model the jet locally as a conducting liquid cylinder surrounded by a gas of permittivity $\varepsilon_0$. The unperturbed jet has a radius $a(z)$, axial velocity $U(z)$, surface tension $\gamma$, and the local normal electric field on its surface $E_s(z)$. Because the liquid is incompressible, the jet radius and velocity are connected by the first approximation as

$$a(z) = \left[\frac{Q}{\pi U(z)}\right]^{1/2}, \quad (1)$$

where $Q$ is the liquid flow rate. Eq. (1) can be used when the velocity is known; experimentally, it is preferable to measure $a$ directly from the image immediately upstream of the transition.

A general small deformation of the jet cross-section can be written as a sum of azimuthal modes[11],

$$r(\theta, z, t) = a(z) + \sum_{m=0}^{\infty} A_m(z,t)\cos[m\theta + \phi_m], \tag{2}$$

where $m$ is the azimuthal mode number, $A_m$ is the time-dependent amplitude of the azimuthal mode $m$, and $\phi_m$ sets the orientation of the deformation. To examine one mode, we write

$$r(\theta, t) = a + \eta(\theta, t), \quad \eta = A_m(t)\cos(m\theta), \quad |A_m| \ll a. \tag{3}$$

The present calculation is local in $z$: axial variation is assumed to be slow compared with variation around the circumference. It therefore determines which non-axisymmetric cross-sectional mode first becomes unstable, rather than providing a complete axial dispersion relation. The value of the azimuthal mode number gives the jet instability[18] (see Figure 5); 0 – Rayleigh varicose breakup, 1 – asymmetric jet perturbations (whipping onset), and 2 – elliptic deformation (sheet formation).

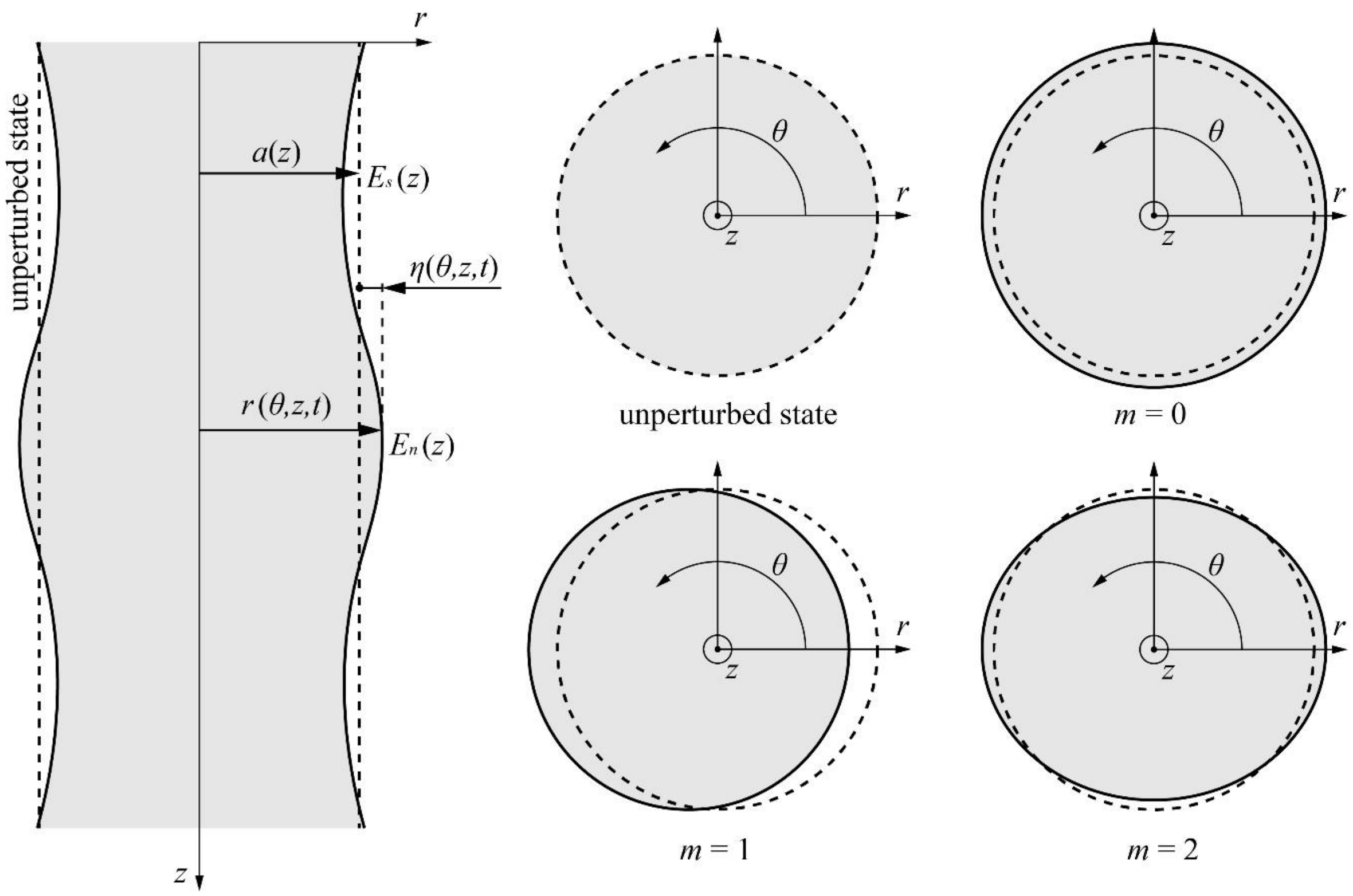

Figure 5: Geometry of the analytical jet representation.

We are primarily interested in the electric sheet onset condition, or in other words, when does a perturbation of the jet diameter grow with the $m = 2$ mode instability. At the outward bulge, which would initiate the sheet, we define the electric-pressure perturbation $\delta p_e$, which promotes further deformation, and the capillary-pressure perturbation $\delta p_\gamma$ which resists it. The net linear driving pressure is therefore

$$\delta p_{\text{net}} = \delta p_e - \delta p_\gamma. \tag{4}$$

When $\delta p_{\text{net}} > 0$, the perturbation grows in one of the aforementioned modes.

*3.1 Capillary restoring pressure*

The capillary restoring contribution comes from the perturbation of the Laplace pressure, $\delta p_\gamma = \gamma\delta\kappa$, where the standard expression for the curvature of a planar polar curve[19] $r = r(\theta)$ is

$$\kappa = \frac{r^2 + 2r_\theta^2 - r r_{\theta\theta}}{\left(r^2 + r_\theta^2\right)^{3/2}}, \tag{5}$$

and $r_\theta$ denotes the first, and $r_{\theta\theta}$ the second derivative of the perturbation with respect to theta, respectively. Substituting $r = a + \eta$ and retaining only linear terms in $a$ and $\eta$ gives,

$$\kappa \simeq \frac{a^2 + 2a\eta - a\eta_{\theta\theta}}{(a^2 + 2a\eta)^{\frac{3}{2}}} = \frac{a^2\left(1 + \frac{2\eta - \eta_{\theta\theta}}{a}\right)}{a^3\left(1 + \frac{2\eta}{a}\right)^{\frac{3}{2}}}. \tag{6}$$

Using the first-order binomial expansion $(1 + x)^{3/2} \simeq 1 + 3x/2$,

$$\kappa \simeq \frac{1}{a}\frac{1 + \frac{2\eta - \eta_{\theta\theta}}{a}}{1 + \frac{3\eta}{a}}, \tag{7}$$

and first-order approximation $1/(1 + x) \simeq 1 - x$, Eq. (6) becomes

$$\kappa \simeq \frac{1}{a}\left[1 + \frac{2\eta - \eta_{\theta\theta}}{a}\right]\left[1 - \frac{3\eta}{a}\right]. \tag{8}$$

Multiplying and again retaining only first-order terms, we derive the curvature of the perturbation $r$

$$\kappa \simeq \frac{1}{a} - \frac{\eta + \eta_{\theta\theta}}{a^2}, \tag{9}$$

where the first term $1/a$ is the curvature of the original circular jet. The second term is the first-order change in curvature $\delta\kappa$ caused by the disturbance. This result is also the $k = 0$ limit of the curvature variation for a disturbed cylindrical surface, as derived by Petrin[11]. For the mode $\eta = A_m\cos(m\theta)$, the second derivative is $\eta_{\theta\theta} = -m^2\eta$, so that

$$\kappa \simeq \kappa_0 + \delta\kappa = \frac{1}{a} + \frac{m^2 - 1}{a^2}\eta. \tag{10}$$

The corresponding perturbation of the Laplace pressure is therefore

$$\delta p_\gamma = \gamma\delta\kappa = \frac{\gamma}{a^2}(m^2 - 1)\eta. \tag{11}$$

*3.2 Electrostatic destabilizing pressure*

Assume that the conducting jet is an equipotential surface. Let $\Phi_0 = \Phi_0(r)$ denote the local base-state electric potential and define the electric field at the unperturbed surface (see Figure 5) as

$$E_s = -\frac{\partial\Phi_0}{\partial r}\bigg|_{r=a}. \tag{12}$$

When the surface is deformed, the original potential $\Phi_0(r)$ no longer satisfies the equipotential boundary condition on the new surface. The total potential is therefore written as

$$\Phi(r,\theta) = \Phi_0(r) + \Phi_1(r,\theta), \tag{13}$$

where $\Phi_1(r,\theta)$ is the small change in potential caused by the surface deformation, and is on the same relatively small order as $\eta/a$. The conducting jet remains at potential $V_j$. Therefore, on the actual surface $r = a + \eta$, $\Phi(a+\eta,\theta) = V_j$, which is already satisfied in the base state $\Phi_0(a) = V_j$. Using $\Phi = \Phi_0 + \Phi_1$, the exact boundary condition is

$$\Phi_0(a+\eta) + \Phi_1(a+\eta,\theta) = \Phi_0(a). \tag{14}$$

We now expand both terms about $r = a$. For the base potential,

$$\Phi_0(a+\eta) = \Phi_0(a) + \eta \left.\frac{d\Phi_0}{dr}\right|_{r=a} + \cdots, \tag{15}$$

and for the perturbed potential,

$$\Phi_1(a+\eta,\theta) = \Phi_1(a,\theta) + \eta \left.\frac{\partial\Phi_1}{\partial r}\right|_{r=a} + \cdots. \tag{16}$$

Because $\Phi_1 = O(\eta)$, its radial derivative is also first order. Consequently, the product $\eta(\partial\Phi_1)/\partial r = O(\eta^2)$ is of second order and is neglected in the linear theory, as with the derivation of the capillary restoring pressure. Substituting Eq. (15) and Eq. (16) into Eq. (14),

$$\Phi_0(a) + \eta \left.\frac{d\Phi_0}{dr}\right|_{r=a} + \Phi_1(a,\theta) = \Phi_0(a), \tag{17}$$

and cancelling $\Phi_0(a)$ gives

$$\Phi_1(a,\theta) = -\eta \left.\frac{d\Phi_0}{dr}\right|_{r=a} = \eta E_s = E_s A_m \cos(m\theta). \tag{18}$$

The displacement of the surface changes the base potential by $\eta\ d\Phi_0/dr$. The perturbation potential $\Phi_1$ must cancel this change so that the conductor remains equipotential. For a remote counter-electrode, the local solution (see Appendix A) of Laplace's equation that decreases away from the jet is

$$\Phi_1(r,\theta) = E_s A_m \left(\frac{a}{r}\right)^m \cos(m\theta). \tag{19}$$

The normal electric field at the perturbed interface $E_n(r)$ is equal to the base electric field $E_0(r)$ and the contribution from the perturbed electric field $E_1(r)$, $E_n \simeq E_0 + E_1$. The base electric field at the displaced surface (for derivation see Appendix B) decreases geometrically as

$$E_0(a+\eta) \simeq E_s\left(1 - \frac{\eta}{a}\right), \tag{20}$$

while the perturbation field contributes (see Appendix B)

$$E_1(a+\eta) \simeq mE_s\frac{\eta}{a}. \tag{21}$$

Consequently, the total normal field at the perturbed interface is

$$E_n \simeq E_s\left[1 + (m-1)\frac{\eta}{a}\right]. \tag{22}$$

The normal field perturbation,

$$\delta E_n \simeq E_n - E_s = E_s(m-1)\frac{\eta}{a}, \tag{23}$$

contains the factor $m - 1$ with two contributions; charge focusing caused by shape $(m)$ and the geometrical decrease of the base field (1). For $m = 1$ (whipping instability) the normal field perturbation is zero, which is consistent with the fact that this mode is a translation of the circular cross-section. For $m = 2$ (elliptic instability), the normal field perturbation scales as $\delta E_n \simeq E_s\eta/a$. An outward bulge, therefore experiences a larger electric field when $m > 1$.

The normal Maxwell pressure on a conducting surface is

$$p_e = \frac{1}{2}\varepsilon_0 E_n^2, \tag{24}$$

while the base Maxwell pressure is defined using Eq. (12),

$$p_{e,0} = \frac{1}{2}\varepsilon_0 E_s^2. \tag{25}$$

A small perturbation changes the base field as $E_n = E_s + \delta E_n$, so that

$$p_e = \frac{1}{2}\varepsilon_0(E_s^2 + 2E_s\delta E_n + \delta E_n^2). \tag{26}$$

The term $\delta E_n^2$ is second order and is neglected, so that

$$p_e \simeq \frac{1}{2}\varepsilon_0 E_s^2 + \varepsilon_0 E_s \delta E_n = p_{e,0} + \delta p_e. \tag{27}$$

We can consequently write the first-order electrical pressure perturbation as

$$\delta p_e = \varepsilon_0 E_s^2 (m-1)\frac{\eta}{a}. \tag{28}$$

At an outward bulge, $\eta > 0$, for $m > 1$, Eq. (28) gives $\delta p > 0$. The bulge, therefore, experiences greater outward electric pressure, causing it to grow further. This positive feedback is why the Maxwell-pressure perturbation has a destabilizing effect.

*3.3 Onset condition*

We evaluate Eq. (4) by substituting Eq. (11) and Eq. (28),

$$p_{\text{net}} = \delta p_e - \delta p_\gamma = \delta p_{\text{net}} = \left[\frac{\varepsilon_0 E_s^2}{a}(m-1) - \frac{\gamma}{a^2}(m^2-1)\right]\eta. \tag{29}$$

Using $m^2 - 1 = (m-1)(m+1)$ and defining the electric Bond number, $\mathrm{Bo_e} = \varepsilon_0 E_s^2 a/\gamma$,

$$\delta p_{\text{net}} = \frac{\gamma}{a^2}(m-1)[\mathrm{Bo_e} - (m+1)]\eta, \tag{30}$$

we find that for $m \geq 2$, the disturbance grows when the quantity in square brackets is positive. The onset condition for the formation of the electric sheet, $m = 2$, is then

$$\mathrm{Bo_e} = \frac{\varepsilon_0 E_s^2 a}{\gamma} = 3. \tag{31}$$

The only unknown left in the presented model is the electric field on the jet surface $E_s$. The applied electric field in the domain, $V/H$, is locally amplified due to the effects of the capillary, plate, and local jet geometry. Coelho and Debeau[20] have previously researched the electrostatic properties of the point-plane electrode configuration in the hyperbolic approximation for the use in dielectric conduction and breakdown, field emission and field ion microscopy and their solution has also been adopted in electrospray configuration[8–10]. The electric field intensity of the point-plate configuration[20] is

$$E = \frac{2V}{r_o \ln \frac{4H}{r_o}}, \tag{32}$$

where $r_o$ is the outer radius of the issuing and charged capillary. We derive the onset voltage by substituting the above equation into Eq. (31),

$$V_c = \frac{r_o}{2} \ln\left(\frac{4H}{r_o}\right)\left(\frac{3\gamma}{\varepsilon_0 a}\right)^{1/2}. \tag{33}$$

*3.4 Comparison with the experiment*

Measuring the jet thickness required in Eq. (33) immediately before sheet formation is not always possible with the current setup because the sheet occasionally forms directly at the nozzle orifice, leaving no optically resolvable section of the undeformed jet. Nevertheless, a limited number of measurements were obtained near the onset threshold under conditions in which the precursor jet remained clearly visible.

Figure 6 compares the onset law, Eq. (33), for different liquid flow rates for both ethanol and isopropanol. The liquid flow rate was non-dimensionlized using the characteristic flow rate $Q_0 = (\gamma^4 \rho_l^{-1} \Delta\psi^3)$^(1/2) proposed by Ganan-Calvo et al.[16] for the EFF jet. The *total available potential energy* per unit volume $\Delta\psi = I_0 V_0/Q + \Delta p_g$ is calculated via the total gas pressure drop across the nozzle $\Delta p_g$, the characteristic electrical current transported by the jet $I_0 = (\gamma K Q)$^(1/2) and the characteristic voltage drop across the cone-to-jet transition region $V_0 = (\gamma L_t \varepsilon_0^{-1})$^(1/2). Here $L_t = Q(K\rho_l^2 \varepsilon_0^{-1}\gamma^{-2})$^(1/3) is the cone-to-jet transition length.

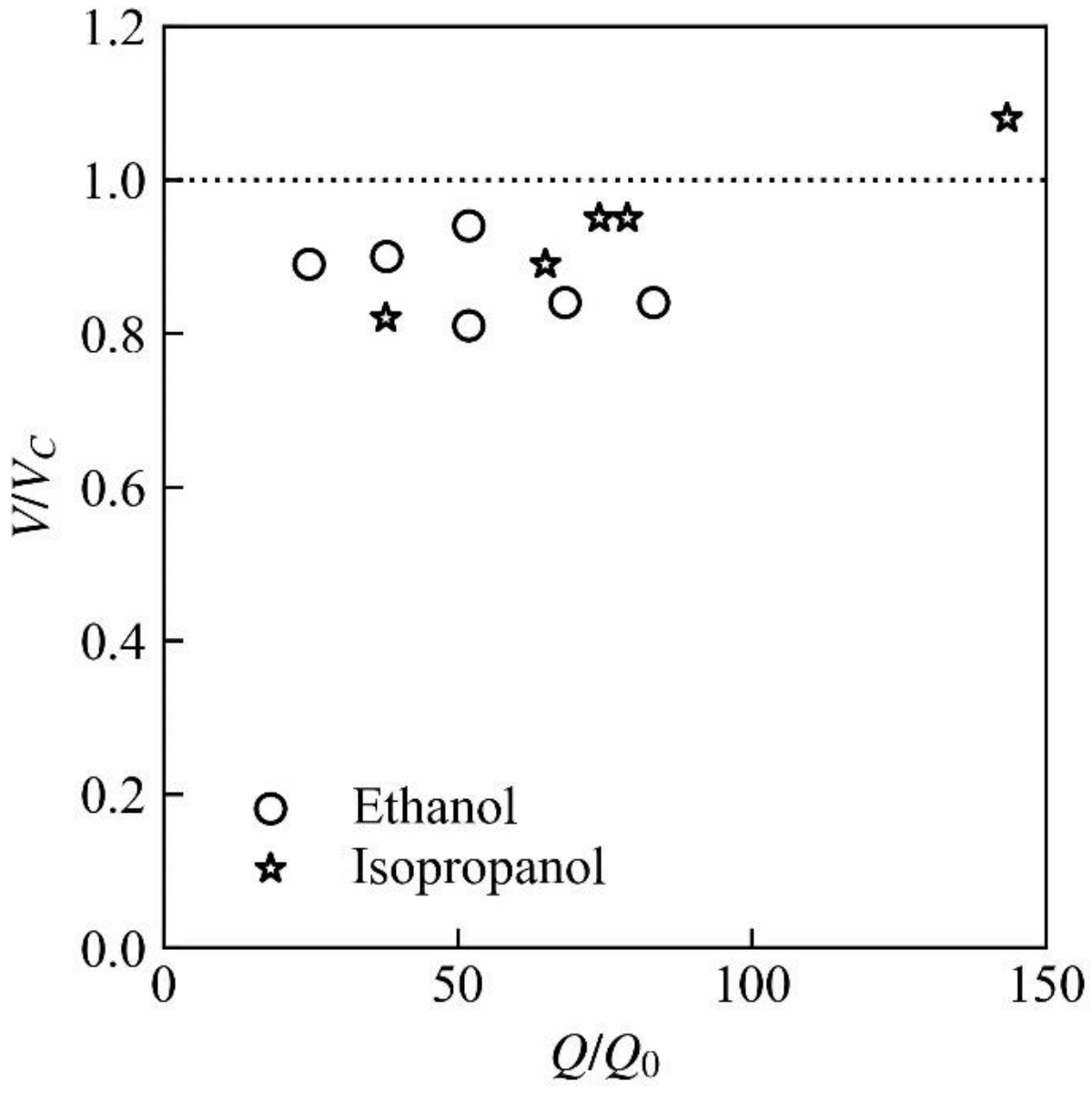

Figure 6: Comparison of the measured onset voltage $V$ with Eq. (33) for different non-dimensional flow rates.

The onset law agrees with our experimental results within the 20% error margin, though it tends to overpredict the onset voltage. The apparent dependence of $V/V_c$ on liquid flow rate, particularly for isopropanol, may reflect hydrodynamic effects absent from the onset criterion, which balances electric and capillary stresses without explicitly accounting for flow. Changes in flow rate could modify the local jet geometry and surface electric field, or the time available for disturbances to grow, thereby shifting the observed onset voltage. However, the limited number of measurements and their scatter prevent a firm conclusion, and the apparent trend may simply be a statistical artefact rather than evidence of a systematic flow-rate dependence.

Interestingly, the onset law does not depend on the jet-generation method and is therefore also generally applicable to cone-jet electrosprays. Wang et al.[10], who used the same electric Bond number definition as defined here, reported the onset of the ramified jet regime for cone-jet electrosprays in the range of $3.0 < \mathrm{Bo_e} < 3.8$ (see Figure 8 in their paper), validating the prediction of the current model.

## 4 Conclusion

An applied electric field transformed a gas-focused jet into a quasi-stable liquid sheet that emitted fine jets from its rims. Increasing voltage preserved the overall sheet topology but reduced its thickness (as seen via optical attenuation), produced more numerous and finer rim jets, and narrowed the spray toward the grounded electrode. The non-monotonic downstream motion indicates that voltage alone does not determine the instantaneous dynamics. Higher gas flow moved the jet-to-sheet transition toward the nozzle and produced finer structures, whereas higher liquid flow produced a longer, wider, and optically denser sheet with thicker rims and a stronger central jet. The irregular ligaments observed at the lowest gas and highest liquid flow are consistent with competition between aerodynamic focusing and liquid inertia. The rim jets grew downstream of the rim and underwent capillary breakup. Finer jets near the beginning of the rim produced smaller droplets, initially emitted transverse to the main flow, while droplet size increased toward the centerline. Their stronger deflection toward the electrode is consistent with a larger charge-to-mass ratio.

A developed sheet formed with ethanol, but not with the higher-surface-tension water or ethylene glycol, while the more viscous isopropanol showed only initial sheet formation. These observations are consistent with capillary and viscous resistance opposing deformation. It would be very interesting to repeat the experiments at a lower scale, as it is normally used in sample delivery systems for crystallography[6], where the interface force balance favors surface tension more strongly.

The analytical model assumed an incompressible, locally circular and slowly varying jet; constant surface tension; an equipotential conducting liquid; a charge-free dielectric gas governed by quasi-static Laplace electrostatics; a remote counter-electrode represented by a locally radial field; and a small, linear azimuthal disturbance independent of $z$. It neglects finite-charge relaxation and transport, space charge, viscous, inertial, and aerodynamic perturbations, axial modes, and nonlinear breakup. Electric properties of ethanol and isopropanol imply a charge-relaxation time $\varepsilon/K$ on the order of $10^{-2}$ s, which is much longer than the observed deformation timescale (<1 ms) and thus justifying the equipotential assumption. Furthermore, the cylindrical boundary approximation assumed in Appendix B for the pre-onset jet is supported in Figs. 2,3 and 4, where the rim jets initially emerge normal to the liquid interface and curve toward the counter-electrode only farther downstream. The model predicts the sheet onset voltage with below 20% error. The onset condition of $\mathrm{Bo_e} = 3$ is not confined to EFF but is generally applicable to electrified liquid jets, for example, the cone-jet electrospray ramified regime reported by Wang et al.[10]. Further studies should examine the influence of liquid flow rate, as both our model's accuracy and previously reported results[10] exhibit a moderate dependence on this parameter.

## Acknowledgments

The funding for this research was provided by the Slovenian Grant and Innovation Agency (ARIS) within Program Group P2-0162.

## Author declarations

### Conflict of Interest

The authors have no conflicts to disclose.

### Author Contributions

**Bor Zupan**: Conceptualization (equal); Methodology (lead); Data Curation (lead); Formal Analysis (lead); Investigation (equal); Visualization(lead); Writing – original draft (lead), Writing – review & editing (equal). **Jernej Kušar**: Conceptualization (equal); Investigation (equal); Resources (lead); Writing – review & editing (equal). **Božidar Šarler**: Conceptualization (equal); Supervision (lead); Funding Acquisition (lead); Supervision (lead); Writing – review & editing (equal).

### Data availability

The data that support the findings of this study are available within the article.

## Appendix A: The perturbation potential

For a two-dimensional disturbance, Laplace's equation $\nabla^2\Phi_1 = 0$ in polar coordinates is

$$\frac{\partial^2\Phi_1}{\partial r^2} + \frac{1}{r}\frac{\partial\Phi_1}{\partial r} + \frac{1}{r^2}\frac{\partial^2\Phi_1}{\partial\theta^2} = 0. \tag{A1}$$

The boundary condition contains $\cos(m\theta)$, so we assume a separated solution of the form $\Phi_1(r,\theta) = R_m(r)\cos(m\theta)$,

$$R_m''(r)\cos(m\theta) + \frac{1}{r}R_m'(r)\cos(m\theta) - \frac{m^2}{r^2}R_m(r)\cos(m\theta) = 0, \tag{A2}$$

where $R_m(r)$ is an unknown radial function. For the equation

$$R_m'' + \frac{1}{r}R_m' - \frac{m^2}{r^2}R_m = 0 \tag{A3}$$

we can try the power-law solution, $R_m(r) = r^n$, so that $R_m' = nr^{n-1}$ and $R_m'' = n(n-1)r^{n-2}$. Substituting this into Eq. (A3) and collecting same order terms gives

$$n(n-1) + n - m^2 = n^2 - m^2 = 0, \tag{A4}$$

therefore,

$$n = \pm m. \tag{A5}$$

The general radial solution is

$$R_m(r) = C_m r^m + D_m r^{-m}. \tag{A6}$$

The term $r^m$ increases without bound as $r$ increases. For a remote counter-electrode, the disturbance caused by jet deformation must decrease with increasing distance from the jet. We therefore set $C_m = 0$. The general solution for the perturbation potential is thus

$$\Phi_1(r,\theta) = D_m r^{-m}\cos(m\theta)\,. \tag{A7}$$

At $r = a$, Eq. (A7) requires $D_m a^{-m}\cos(m\theta) = E_s A_m\cos(m\theta)$, therefore

$$D_m = E_s A_m a^m. \tag{A8}$$

The perturbation potential is consequently

$$\Phi_1(r,\theta) = E_s A_m \left(\frac{a}{r}\right)^m \cos(m\theta). \tag{A9}$$

## Appendix B: Geometrical reduction of the base and perturbed fields

### Contribution from the base field

Assume no free charges in the gas, so that the Laplace's equation $\nabla^2\Phi_0 = 0$ holds. For an axisymmetric potential with no $z$-dependence,

$$\frac{1}{r}\frac{d}{dr}\left(r\frac{d\Phi_0}{dr}\right) = 0, \tag{B1}$$

the solution is $\Phi_0(r) = C_1 \ln r + C_2$.. Let us approximate the remote counter-electrode by a cylindrical boundary at $r = b$, where $b \gg a$. The boundary conditions are $\Phi_0(a) = V_j, \Phi_0(b) = 0$. Solving for constants $C_1, C_2$ gives

$$\Phi_0(r) = V_j \frac{\ln(b/r)}{\ln(b/a)} \tag{B2}$$

The base electric field $\boldsymbol{E}_0$ depends only on $r$, so that

$$\boldsymbol{E}_0(r) = -\nabla\Phi_0 = \mathbf{E}_0 = -\frac{d\Phi_0}{dr}\hat{\mathbf{e}}_r. \tag{B3}$$

Differentiating Eq. (B2), and omitting the vector notation, we can rewrite Eq. (B3) as

$$E_0(r) = \frac{V_j}{r\ln(b/a)}. \tag{B4}$$

For coaxial approximation, the surface electric field $E_s$ (see Eq. (12)) is then

$$E_s = \frac{V_j}{a\ln\left(\frac{b}{a}\right)}, \tag{B5}$$

and the base electric field can be written as

$$E_0(r) = E_s\frac{a}{r}. \tag{B6}$$

In the real experiment, the counter-electrode is a plate rather than a cylindrical electrode. Eq. (B6) is therefore a local cylindrical approximation. The actual global geometry is contained in the measured or calculated value of $E_s$.

For the perturbed jet at $r = a + \eta$, the electric field is

$$E_0(a+\eta) = E_s\frac{a}{a+\eta} = E_s\frac{1}{1+\eta/a}. \tag{B7}$$

Using $1/(1+x) \simeq 1 - x$,

$$E_0(a+\eta) \simeq E_s\left(1 - \frac{\eta}{a}\right), \tag{B8}$$

we derived the geometrical reduction of the base field because the outward bulge is farther from the jet axis.

**Contribution from the perturbed field**

The perturbed electric field,

$$E_1 = -\frac{\partial \Phi_1}{\partial r}\bigg|_{r=a}, \tag{B9}$$

is calculated from Eq. 19, as

$$\frac{\partial \Phi_1}{\partial r} = -mE_s A_m a^m r^{-m-1} \cos(m\theta). \tag{B10}$$

We will evaluate this term at $r = a$, not $a + \eta$, because $\partial \Phi_1/\partial r$ is already first order. Correcting its evaluation position would produce a second-order term.

$$\frac{\partial \Phi_1}{\partial r}\bigg|_{r=a} = -mE_s A_m \frac{a^m}{a^{m+1}} \cos(m\theta) = -mE_s \frac{A_m}{a} \cos(m\theta). \tag{B11}$$

Using $\eta = A_m \cos(m\theta)$, we get

$$E_1 = -\frac{\partial \Phi_1}{\partial r}\bigg|_{r=a} = mE_s \frac{\eta}{a}. \tag{B12}$$